\documentclass[a4paper]{article}

\usepackage{INTERSPEECH2021}
\usepackage{comment}
\usepackage{multirow}
\usepackage{color}
\usepackage[ruled]{algorithm}
\usepackage{algpseudocode}
\usepackage{caption} 
\usepackage{cite} 
\usepackage{subcaption}
\title{Streaming End-to-End ASR based on Blockwise Non-Autoregressive Models}
\name{Tianzi Wang$^1$, Yuya Fujita$^2$, Xuankai Chang$^{1,3}$, Shinji Watanabe$^{1,3}$}
\address{
  $^1$Johns Hopkins University\\
  $^2$Yahoo Japan Corporation\\
  $^3$Carnegie Mellon University}
\email{wtianzi1@jhu.edu, yuyfujit@yahoo-corp.jp, \{xuankaic,swatanab\}@andrew.cmu.edu}

\begin{document}

\maketitle
\begin{abstract}
Non-autoregressive (NAR) modeling has gained more and more attention in speech processing. With recent state-of-the-art attention-based automatic speech recognition (ASR) structure, NAR can realize promising real-time factor (RTF) improvement with only small degradation of accuracy compared to the autoregressive (AR) models. However, the recognition inference needs to wait for the completion of a full speech utterance, which limits their applications on low latency scenarios. To address this issue, we propose a novel end-to-end streaming NAR speech recognition system by combining blockwise-attention and connectionist temporal classification with mask-predict (Mask-CTC) NAR. During inference, the input audio is separated into small blocks and then processed in a blockwise streaming way. To address the insertion and deletion error at the edge of the output of each block, we apply an overlapping decoding strategy with a dynamic mapping trick that can produce more coherent sentences. Experimental results show that the proposed method improves online ASR recognition in low latency conditions compared to vanilla Mask-CTC. Moreover, it can achieve a much faster inference speed compared to the AR attention-based models. All of our codes will be publicly available at https://github.com/espnet/espnet.


\end{abstract}

\section{Introduction}
\label{sec:introduction}

Over the past years, the advances in deep learning have dramatically boosted the performance of end-to-end (E2E) automatic speech recognition (ASR) \cite{graves2006connectionist, graves2012sequence, chorowski2014end}. Most of these E2E-ASR studies were based on autoregressive (AR) models and they achieved state-of-the-art performance \cite{gulati2020Conformer}. However, there is a disadvantage of the AR models in that the inference time linearly increases with the output length. Recently, non-autoregressive (NAR) models has gained more and more attention in sequence-to-sequence tasks, including machine translation \cite{gu2017non,libovicky2018end,ghazvininejad2019mask}, speech recognition (ASR) \cite{graves2006connectionist,chan2020imputer,fujita2020insertion,chi2020align,fan2020cass}, and speech translation \cite{inaguma2020orthros}.
In contrast to the AR modeling, NAR modeling can predict the output tokens concurrently, the inference speed of which is dramatically faster than AR modeling. 
Especially, connectionist temporal classification (CTC) is a popular and simple NAR modeling \cite{graves2006connectionist,libovicky2018end}. 
However, CTC makes a strong conditional independent assumptions between the predicted tokens, leading to an inferior performance compared to the AR attention-based models \cite{battenberg2017exploring}. 

To overcome this issue, several NAR studies have been proposed in the ASR field.
A-FMLM\cite{chen2019listen} is designed to predict the masked tokens conditioning on the unmasked ones and the input speech. However, it needs to first predict the output length, which is difficult and easily leads to a long output sequence.
Imputer \cite{chan2020imputer} directly used the length of input feature sequence to address the issue and achieves comparable performance with AR models, but the computational cost can be very large. 
ST-NAR \cite{tian2020spike} used CTC to predict the target length and to guide the decoding. It is fast but suffers from large accuracy degradation compared with AR models.
Different from previous methods, Mask-CTC \cite{higuchi2020mask} first generates the output tokens with greedy decode of a CTC, and then refines the tokens which have low confidence by a mask-predict decoder \cite{ghazvininejad2019mask}. 
Mask-CTC usually predicts sequences with reasonable length and can achieve a fast inference speed, 7x faster than the AR attention-based model.

In addition to fast inference, latency is an important factor to be considered for the ASR system used in a real-time speech interface. 
There have been a lot of prior studies for low-latency E2E ASR based on Recurrent Neural Networks Transducer (RNN-T) \cite{graves2012sequence,battenberg2017exploring,tripathi2020transformer,jain2019rnn} and online attention-based encoder decoder (AED) with AR models, such as monotonic chunkwise attention (MoChA) \cite{chiu2017monotonic, inaguma2020enhancing}, triggered attention \cite{moritz2020streaming}, and blockwise-attention \cite{miao2020transformer,tsunoo2020streaming}.

Motivated by such online AR AED studies and emergent NAR research trends, we propose a novel end-to-end streaming NAR speech recognition system, by combining the blockwise-attention and Mask-CTC models. 
During inference, the input audio is first separated into small blocks with $50\%$ overlap between consecutive blocks. CTC firstly predicts the preliminary tokens per block with an efficient greedy forward pass based on the output of a blockwise-attention encoder. To address the insertion and deletion error of CTC outputs frequently appeared at the boundary of each block, we apply a dynamic overlapping strategy \cite{chiu2019comparison} to produce coherent sentences. Then, low-confidence tokens are masked and re-predicted by a mask-predict NAR decoder \cite{higuchi2020improved} conditioning on the rest tokens. The greedy CTC, dynamic overlapping decoding, and mask-prediction all perform very fast, thus can achieve quite low RTF. We evaluated our approach on TEDLIUM2 \cite{rousseau2014enhancing} and AISHELL1 \cite{bu2017aishell}. Compared to vanilla full-attention Mask-CTC, our proposed method decreases the online ASR recognition error rate in a low latency condition, with very fast inference speed. To the best of our knowledge, this is the first work that extending the NAR mechanism into streaming ASR.

\subsection{Relationship with other streaming ASR studies}
The most important success of streaming ASR recently is RNN-T and its variants. RNN-T based systems achieve state-of-the-art ASR performance for streaming applications and are successfully deployed in production systems\cite{tripathi2020transformer,jain2019rnn,li2020towards,mahadeokar2021alignment}. 
However, the recurrent mechanism predicts the token of the current input frame based on all previous tokens using recurrent layers, to which NAR cannot be easily applied. 
Besides, several ideas in this paper are inspired from online AR AED architectures\cite{chiu2019comparison,miao2020transformer,tsunoo2020streaming}, since they have more technical connections with NAR models based on the similar encoder-decoder framework.

\section{Mask-CTC}
\label{sec:mask-ctc}
Mask-CTC is a non-autoregressive model trained with both CTC objective and mask-prediction objective \cite{higuchi2020mask}, where the mask-predict decoder predicts the masked tokens based on CTC output tokens and encoder output. 
CTC predicts a frame-level input-output alignment based on conditional independence assumption between frames. It models the probability $P(Y|X)$ of output sequence by summing up all possible alignments, where $Y$ denotes the sequence of output and $X$ denotes the input audio.
However, due to the conditional independence assumption, CTC loses the ability of modeling correlations between output tokens and consequently loses performance.

Mask-CTC was designed to mitigate this issue by adopting an attention-based decoder as a masked language model (MLM) \cite{ghazvininejad2019mask,chen2019nonautoregressive}, and iterative refining the output of CTC greedy decoding. During training, the tokens in the ground-truth are randomly selected and replaced by a special $\langle\text{mask}\rangle$ token. Then the decoder is trained to predict the actual tokens at the masked positions, $Y_\text{mask}$, conditioning on the rest unmasked tokens, $Y_\text{obs}$, and attention-based encoder output. $\mathbf{H_{\text{enc}}}$.
\begin{align}
\label{eq:encoder}
\mathbf{H_{\text{enc}}}&=\text{MHSAEncoder}(X),\\
P_{\text{MLM}}(Y_{\text{mask}}|Y_{\text{obs}}, \mathbf{H_{\text{enc}}})&=\prod_{y_\text{mask}\in Y_{\text{mask}}}P(y_\text{mask}|Y_{\text{obs}}, \mathbf{H_{\text{enc}}}).
\end{align}
where the $\text{MHSAEncoder}(\cdot)$ denotes a multi-headed self-attention based encoder.
The Mask-CTC model is optimized by a weighted sum of the CTC and MLM objectives:
\begin{align}
\mathcal{L}=&\gamma \log P_{\text{ctc}}(Y|\mathbf{X}) + (1-\gamma) \log P_{\text{MLM}} (Y_{\text{mask}}|Y_{\text{obs}},\mathbf{X}),
\end{align}
where $\gamma$ is a tunable hyper-parameter.

\section{Streaming NAR}
\label{sec:streaming_nar}
The overall architecture of the proposed E2E streaming NAR speech recognition system is shown in Figure 1. The main difference compared with Mask-CTC is that the normal MHSA-based encoder (e.g. Transformer/Conformer) as shown in Eq.~\eqref{eq:encoder} is replaced by a blockwise-attention encoder to make the model streamable.
\begin{figure}[t]
  \centering
  \includegraphics[width=0.9\linewidth]{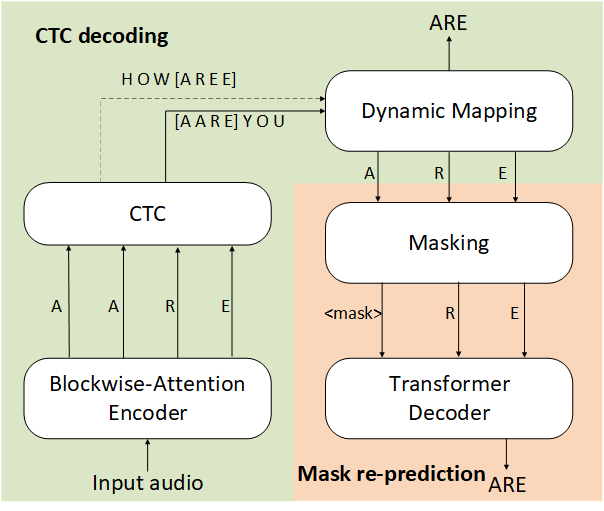}
  \caption{Architecture of proposed streaming NAR}
  \label{fig:s}
\end{figure}
\subsection{blockwise-attention Encoder}
\label{sec:bloc_att_end}
To build a streaming AED-based ASR system, the encoder is only allowed to access limited future context.
We use a blockwise-attention (BA) based encoder \cite{miao2020transformer,tsunoo2020streaming} instead of normal multi-headed self-attention (MHSA).
In a BA based encoder, the input sequence X is divided into fix-length blocks $X=[X_b]_{b=0}^{B}$, where $X_b=[x_{bl}, \dots, x_{(b+1)l-1}]$. Here $b$ is the block index, $B$ is the index of the last block in the whole input and $l$ is the block length. In the computation of blockwise-attention, each block only attends to the former layer's output within the current block and the previous block. Blockwise-attention at the $b$-th block is defined as:
\begin{align}
Z_b^{i}&=\text{BA}(Z_b^{i-1})=\text{MHSA}(Z_b^{i-1}, [Z_{b-1}^{i-1}, Z_{b}^{i-1}], [Z_{b-1}^{i-1}, Z_{b}^{i-1}]),
\label{eq:ba}
\end{align}
where $Z_b^i$ is the output of encoder layer $i$ at $b$-th block, $Z^0=X$. 
The three arguments of $\text{MHSA}(\cdot)$ are query, key, and value matrix variables, respectively.
Likewise, the blockwise-depthwise-convolution (BDC) in Conformer encoder is defined as:
\begin{align}
    \text{BDC}(Z_b^{i-1})=\text{CONV}(\text{BPAD}(Z_b^{i-1},[Z_{b-1}^{i-1}, Z_{b}^{i-1}])), 
    \label{eq:bdc}
\end{align}
where $\text{CONV}(\cdot)$ is 1D depth-wise convolution and $\text{BPAD}(\cdot)$ refers to blockwise-padding, that pads zeros at right edges and pads $Z_{b-1}^{i}$ at left edges of $Z_b^{i-1}$ to keep the input/output dimension identical. The rest operations, such as point-wise convolution and activation function, is the same as in Conformer \cite{guo2020recent}.
\subsection{Blockwise Mask-CTC}
As shown in Figure 1, the proposed E2E streaming NAR speech recognition system consists of a blockwise-attention based encoder, a CTC, a dynamic mapping process, and an MLM decoder. During training, we use full audio as the input for convenience. But the CTC output within a block only depends on the input in the current and previous block, since the computation of the encoder is blockwise. In this paper, following the NAR manner, we applied greedy decoding for CTC, which selects the token with the highest probability at each time step. The output of each block from greedy decoding CTC is:
\begin{align}
\pi_b&=\text{Concat}_{y_t}[\arg\max_{y_{t_{i,b}}}P(y_{t_{i,b}}| \text{BAEncoder}(X_b))]
\end{align}
where $b\in [0,..,B]$, $t_{i,b}$ denotes the $i$-th time step belonging to the $b$-th block, $t_{i,b} \in [t_{0,b}, \dots, t_{l-1,b}]$. BAEncoder refers to the blockwise-attention encoder as mentioned in Sec \ref{sec:bloc_att_end}, $X_{-1}$ is set to be a zero-matrix with the same size as $X_{b}$ during computation. At the same time, the ground-truth target tokens are randomly masked and re-predicted by the MLM decoder.

During inference, the input is online segmented into fixed-length blocks with 50\% overlap and fed into the encoder in a streaming way. The encoder forward pass and the CTC decoding follow the same way as in the training. As shown in Figure \ref{fig:s}, the dotted line denotes CTC output $y_{b-1}$ corresponds to input block $X_{b-1}$, and solid line denotes CTC output $y_b$ of $X_b$. A dynamic mapping trick is applied on the overlap tokens between $y_{b-1}$ and $y_{b}$ to make a coherent output. More details will be shown in Sec. \ref{sec:dy}. Then the tokens with low-confidence scores from CTC decoding outputs, $\pi_{0..B}$, are masked and re-predicted by MLM decoder. The predicting process is done in several iterations. In each iteration, tokens with higher predicting probability are filled into masked positions and the re-filled sequences are used as input for the next iteration.
\begin{align}
\hat{Y}_0&=\text{mask}(\pi_b)\\
y_{k,m} &= \arg\max\log P_\text{MLM}(y_{mask}|\hat{Y}_{m-1},X)
\end{align}
where $\hat{Y_0}$ denotes the masked token sequence from CTC output, $m$ denotes the iteration number of re-prediction, $y_{k,m}$ denotes $k$ re-predicted tokens in $m_{th}$ iteration, $y_{k,m}$ are then infilled into the predicted sequences $\hat{Y}_{m-1}$ at corresponding masked positions to form the $m_{th}$ predition $\hat{Y}_{m}$. $\hat{Y}_N$ would be the final output as $N$ is total number of iterations.
\begin{table*}[tb]
  \caption{TEDLIUM2: WERs on dev/test, averaged Latency and RTF are reported. 640ms input segments is used for all streaming decode mode. The attention-based AR and Mask-CTC models trained with conventional attention, while the proposed streaming NAR use blockwise attention with 640ms block length.}
  \label{tab:example}
  \centering
  \begin{tabular}{clcccccc}
    \toprule
    & \textbf{Encoder type} & \textbf{Decode Mode} & \textbf{WER on dev}  & \textbf{WER on test} &  \textbf{Latency$_{\text{utt}}$(ms)} &\textbf{RTF} \\
    \midrule
    & Transformer           & full-context  & 11.7                 & 9.9                 &5220          &0.46\\
    & \quad  + beamsize=10           & full-context  & 10.9                  & 9.1                  &34160        &3.01\\
\multirow{-2}{*}{\textbf{Attention-based AR}}
&  Conformer             & full-context                                 & 11.0                  & 8.4                  &6130               &0.54\\
& \quad + beamsize=10             & full-context                                 & 11.1                  & 8.1                  &37780            &3.33\\
    \midrule
    & Transformer           & full-context   & 11.0                   & 10.7                 &790            &0.07 \\
    & Conformer             & full-context   & 10.0                   & 8.8                  &1070            &0.09  \\
    & Transformer           & streaming      & 18.2                     &16.4                   & 300  & 0.20             \\
\multirow{-4}{*}{\textbf{Mask-CTC}}                       & Conformer             & streaming     
&23.5   &21.3  &310  &0.26              \\
    \midrule
    & Transformer           & full-context   & 12.2                 & 11.2                 &910              &0.08\\
    & Conformer             & full-context   &\textbf{10.4}                 & \textbf{9.4}                  &1030               &0.09\\
    & Transformer           & streaming      & 14.2        & 14.0             &\textbf{ 120} & 0.22\\
\multirow{-4}{*}{\textbf{Streaming NAR (Proposed)}} & Conformer             & streaming    &\textbf{12.1}
&\textbf{11.7}   & \textbf{140}   & 0.32       \\      
    \bottomrule
  \end{tabular}
\end{table*}
\subsection{Dynamic mapping for overlapping inference}
\label{sec:dy}
Although splitting the input audio into small blocks with fixed-length is a straightforward approach to form streaming ASR, it will result in horrible performance degradation at the block boundaries. A segment boundary may appear in the middle of a token, leading to that one token may have repetitive recognition or non-recognition in two consecutive blocks. The VAD-based segmentation method is a way to solve the issue, but it is sensitive to the threshold and may lead to large latency.

In this paper, we applied overlapping inference with dynamic mapping tricks as in\cite{chiu2019comparison} to recover the erroneous output at the boundary, as shown in Algorithm 1. During inference, we use $50\%$ overlap when segmenting the input audio, which ensures any frame of input audio is predicted twice by Encoder and CTC. By locating the token index in $y_b$ that is closest to the center point of $C_b$, we dynamically search for the best alignment between ($y_{b,:idx_b}$ and $y_{b-1,idx_{b-1}:}$). 
Normalize refers to removing repeated and blank token from CTC output $C_b$.
Following the scoring function: $\text{Score}(t_j^b) = -|j-(l-1)/2|$, we select one of the token in token pairs on the best alignment path as the output of the overlapped segment. Here $t_j^b$ refers to the $j$-th token in $b$-th block, and $l$ is block length.
\begin{algorithm}
\caption{CTC overlap decode and dynamic map}
\begin{algorithmic}[1]
    \State $y_{out}$ = [$\varnothing$]\textbf{};
    \State $\mathbf{X}$ = Audio blocks iterator with $50\%$ overlap\;
    \For{$b$ = 0 to B}\;
    \State $C_b$ = CTC\_Predict(BAEncoder($X_b$))\;
    \If{$ b > 0$}
    \State $y_b$ = remove repeated tokens in $C_b$\;
    \State $idx_b$ = token index in $y_b$ that is closest to $C_{b, \frac{1}{2}l}$\;
    \State $\text{path}_{b}$ = Alignment($y_{b,:idx_b}$, $y_{b-1,idx_{b-1}:}$)\;
    \State $y_{out,b} = [\varnothing]$\;
    \For{($p$, $q$) in $\text{path}_b$}\;
    \State token = $p$ if Score($p$)$>$Score($q$) else $q$
    \State APPEND token to $y_{out,b}$\;
    \EndFor
    \State $y_{out,b}$ = Normalize($y_{out,b}$)\;
    \Else
    \State $y_{out,b}$ = Normalize($C_{b,\frac{1}{2}l}$)\;
    \EndIf 
    \State APPEND $y_{b}$ to $y_{out}$\;
    \EndFor
\end{algorithmic}
\end{algorithm}
\section{Experiment}
\subsection{Experimental setup and Dataset}

We evaluate our proposed model on both Chinese and English Speech corpora: TEDLIUM2\cite{rousseau2014enhancing} and AISHELL1\cite{bu2017aishell}. 
For all experiments, the input features are 80-dimensional log-mel filter-banks with pitch computed with frame length of 25ms and frame shift of 10ms. We use Kaldi toolkit\cite{Povey_ASRU2011} for feature extraction. We also apply speed perturbation(speed rate=0.9, 1.0, 1.1) and spectrum augmentation\cite{park2019specaugment} for data augmentation. Models are evaluated with both full-context decoding and streaming decoding.

All experiments are conducted using the open-source, E2E speech processing toolkit ESPnet\cite{watanabe2018espnet,karita2019comparative,guo2020recent}. The encoder first contains 2 CNN blocks to downsample the input to $1/4$, followed by 12 MHSA-based layers. In streaming NAR, BA-based layers(Eq. \ref{eq:ba}, \ref{eq:bdc}) is used in encoder to replace MHSA. Decoders have a similar stacked-block structure with 6 layers and only full-attention transformer blocks. For any self-attention block in this paper, we use $h = 4$ parallel attention heads, with dimension $d^\text{att} = 256$. For feed-forward layer, we use dimensionality $d^\text{ffn} = 2048$ and apply swish as activation functions. In self-attention, relative position embedding is augmented in the input\cite{dai2019transformer}. 
We use $15$ as the kernel size for Conformer convolution. 
Models on TEDLIUM2 are trained for 200 epochs and on AISHELL1 are trained for 150 epochs. The evaluation is done on the averaged model over the best 10 checkpoints. We do not integrate Language Model during decoding.

\begin{table*}[tb]
  \caption{AISHELL1: WERs, averaged latency and RTF are reported. 1280\text{ms} input segments is used for all streaming inference. The attention-based AR and Mask-CTC models work with conventional attention, while the proposed streaming NAR use blockwise attention with 1280ms block length}
  \label{tab:ais}
  \centering
  \begin{tabular}{clcccccc}
    \toprule
    & \textbf{Encoder type} & \textbf{Decode Mode} & \textbf{WER on dev}  & \textbf{WER on test}  &\textbf{Latency$_{\text{utt}}$(ms)} & \textbf{RTF} \\
    \midrule
    & Transformer             & full-context                                 &  6.7                 & 7.6                  & 2040          &0.45\\
     \multirow{-2}{*}{\textbf{Attention-based AR}}    & \quad + beamsize=10           & full-context  & 6.6                 & 7.4          & 11640     &2.56\\                          
    \midrule
    & Transformer             & full-context   &6.9                    & 7.8                  &220       & 0.05  \\
\multirow{-2}{*}{\textbf{Mask-CTC}}     & Transformer             & streaming     
&9.0   & 10.4   & 280              &0.18\\
 
    \midrule
    & Transformer             & full-context   &8.6                  &9.9                   &230         &  0.05\\
\multirow{-2}{*}{\textbf{Streaming NAR (Proposed)}} & Transformer             & streaming    
&\textbf{8.6}   &\textbf{9.9}     & 320    &0.20 \\      
    \bottomrule
  \end{tabular}
\end{table*}
\subsection{Results}
Table 1 shows TEDLIUM2 results, including word error rates (WERs) on dev/test sets, averaged latency and real-time factor (RTF). Latency and RTF were measured on the CPU platform (Intel(R) Xeon(R) CPU E5-2686 v4 @ 2.3GHz) with 8 parallel jobs. We set block length $l=16$, corresponding to $640$ms before subsampling. 
The full-context evaluation is done on the utterance level, while the streaming evaluation is done on the unsegmented streaming audio for each speaker.
Since it is hard to define the latency with long unsegmented audio, we calculated the averaged latency per utterance with dev:
\begin{align}
\begin{split}
\text{Latency}&=\frac{\sum_{utt}(\text{last token emitted}-\text{end of speech})}{\#\text{total number of utterance}}.
\end{split}
\end{align}
The last token emitted time is the time stamp that model predicts last token, and the end of speech is determined by forced alignment with external model. This latency considered both look-ahead latency and computation time.
For comparison, we also report the latency for full-context decoding by measuring the average decoding time per utterance, since in full-context case decoding can not start before entire audio is fed.
\begin{table}[tb]
  \caption{WERs and RTF on TEDLIUM2 conduct on proposed Streaming NAR model with different block length(BL), Experiments are all conducted under streaming mode.}
  \label{tab:tedlen}
  \centering
  \begin{tabular}{cccccc}
    \toprule
    \textbf{Encoder}                 & \textbf{Dev}                          & \textbf{Test}                                 & \textbf{BL(ms)}                & \textbf{$\text{Latency}_{\text{utt}}$(ms)}  & \textbf{RTF} \\ 
    \midrule
      & 12.8                 & 12.0                         & 5120 & 1530                                     &              0.17 \\
     & 12.8                 & 11.9                         & 2560     & 700                                      &               0.18\\
      & 13.0                 & 12.2                         & 1280   & 290                                      &              0.18 \\
\multirow{-4}{*}{\textbf{BA-TF}}   &14.2      & 14.0          &640   &  120                  &                             0.22 \\
\midrule 
      &10.5                   & 10.3                          & 5120    & 1700                                       &              0.29 \\
     &10.6                   & 10.4                          & 2560  &  790                                         &              0.29 \\
    & 11.2                 & 10.7                          & 1280  &  380                                           &              0.30 \\
\multirow{-4}{*}{\textbf{BA-CF}} 
      &\textbf{12.1}                  &\textbf{11.7}       &640    & 140                                   &              0.32 \\          
    \bottomrule 
  \end{tabular}
\end{table}

Compared to the vanilla Mask-CTC, the proposed streaming NAR model performs much better in streaming mode. With the Conformer encoder, the WER on the dev of Mask-CTC is 23.5 with 310ms averaged latency and that of streaming NAR is 12.1 with 140ms averaged latency. The RTF is about 2x/10x faster than the AR attention-based model with beam size=1/10 respectively. Mask-CTC models are trained with full context input and work better with full future information. We can observe a significant WER degradation in streaming decoding mode, which is due to the input mismatch during training and decoding. On the other hand, the proposed streaming NAR can recognize the current frame with only a small future context, thus fit well with the streaming inference. However, the WERs increased from full context to streaming mode in the proposed model (from 10.4/9.4 to 12.1/11.7). The reason would be the error raised at the segment boundaries. From our observation, these errors can be relieved but not totally solved by dynamic mapping and still exist even with large block lengths.

Table \ref{tab:ais} shows the results on AISHELL1. Since the hyper-parameters for Conformer and Mask-CTC need careful tuning and the model is easy to be overfitted in our preliminary experiments, we only report transformer results on the AISHELL1 task. The vanilla Mask-CTC works better when full-context is provided during decoding while streaming NAR is better on streaming decode, but the benefits are smaller than those in TEDLIUM2. A possible reason is that the training utterance in AISHELL1 is much shorter than TEDLIUM2.
Full-context attention can also gain the ability to recognize with only short future context.

To understand the performance of streaming NAR under different latency, in Table \ref{tab:tedlen} we compare the WERs with different block lengths for blockwise-attention Transformer (BA-TF) and blockwise-attention Conformer (BA-CF) on TEDLIUM2. 
We observe that as the block length get shorter, the latency becomes smaller while RTF rises since the shorter blocks require more iterations to forward the whole input audio. 
Besides, when block length decreases from 5120ms to 1280ms, the result rarely changes. It indicates that in streaming ASR, the closer future context is a much more active player than distant ones. 

\section{Conclusion}
In this paper, we proposed a novel E2E streaming NAR speech recognition system. Specifically, we combined the blockwise-attention based Encoder and Mask-CTC. Beside, we applied the dynamic overlapping inference to mitigate the errors at the boundary. Compared to vanilla Mask-CTC, the proposed streaming NAR model achieves competitive performance in full-context decoding and outperforms the vanilla Mask-CTC streaming decoding with very low utterance latency. Moreover, the decoding speed of the proposed model is about 2x/10x faster than the AR attention-based model with beam size=1/10 respectively. Our future plan is developing better boundaries localization method to replace the overlapping inference, and integrating external language model during decoding.
\section{Acknowledgements}
We would like to thank Mr. Yusuke Higuch of Waseda University for his valuable
information about the Mask-CTC.
\bibliographystyle{IEEEtran}

\bibliography{mybib}

\end{document}